\documentclass[a4paper]{jpconf}
\usepackage{graphicx}
\begin{document}
\newcommand{\RBMO}{$R$BaMn$_{2}$O$_{6}$}

\title{$R$-site Randomness Effect in Double-Perovskite \RBMO}

\author{Y. Miyauchi, S. Fukushima, J. Tozawa, M. Akaki, H. Kuwahara, and D. Akahoshi}

\address{Department of Physics, Sophia University, Tokyo 102-8554, Japan}

\ead{y-miyauc@sophia.ac.jp}

\begin{abstract}
We have investigated the $R$-site randomness effect of $R$/Ba-ordered $R$BaMn$_{2}$O$_{6}$ ($R$ = rare earth) by using Y$_{1-y}$La$_y$BaMn$_2$O$_6$ (0 \(\leq\) $y$ \(\leq\) 1) in which $R$ (Y,La) and Ba are regularly arranged while Y and La randomly occupy the $R$-site.
YBaMn$_2$O$_6$ ($y$ = 0) undergoes charge/orbital ordering (CO) transition at $T_{\rm CO}$ = 500~K while LaBaMn$_{2}$O$_{6}$ ($y$ = 1) shows ferromagnetic metallic (FM) behavior below 350~K\@.
In 0 \(\leq\) $y$ \(\leq\) 0.6, $T_{\rm CO}$ decreases with an increase in $y$.
In 0.6 \(\leq\) $y$ \(\leq\) 0.8, $R$-site randomness causes strong phase separation tendencies among the FM, antiferromagnetic (AFM), and CO insulating (COI) states.
The large magnetoresistance is observed in the phase separated region.
\end{abstract}

\section{Introduction}
Transition-metal oxides with perovskite structure and their derivatives have been intensively studied in terms of technological applications as well as fundamental physics because of their rich physical properties such as high-$T_{\rm C}$ superconductivity in Cu oxides and the colossal magnetoresistance (CMR) effect in Mn oxides[1]\@. 
In conventional perovskites, $R$ and $Ae$ atoms randomly occupy the \mbox{$A$-site} of perovskite structure. 
Resultant $A$-site randomness reduces the onset temperature of the electronic phases, which makes its practical application difficult. 
Therefore, it is required to reduce or remove such $A$-site randomness for realizing its practical application.

Perovskite oxides with $A$-site ordered structures, which are free from $A$-site randomness, are promising candidates for electronic devices using their functionality.
Owing to $A$-site cation ordering, the electronic phase transition temperatures of $A$-site ordered perovskites sometimes exceed room temperature.
For example, $A$-site ordered $R$BaMn$_{2}$O$_{6}$ ($R$ = Y-Sm) is reported to undergo charge/orbital ordering transition at $T_{\rm CO}$ = 500-380~K[2,3]\@.
In the electronic phase diagram for $A$-site ordered $R$BaMn$_2$O$_6$, the ferromagnetic metallic (FM), charge/orbital ordered insulating (COI), and $A$-type antiferromagnetic (AFM) phases compete with each other in a multicritical manner at $R$ = Nd[3]\@.
Mixture of $R$ and Ba, i.e., $A$-site disordering considerably suppresses these ordered phases. 
Consequently, large phase fluctuation is induced near the multicritical point of $R$ = Nd. 
Such large phase fluctuation is significant for the CMR effect[4,5]\@.
In this study, in order to clarify the $R$-site randomness effect on $A$-site ordered $R$BaMn$_{2}$O$_{6}$, we have investigated the magnetic and transport properties of Y$_{1-y}$La$_y$BaMn$_2$O$_6$ (0 \(\leq\) $y$ \(\leq\) 1) in which $R$ (Y,La) and Ba are regularly arranged while Y and La randomly occupy the $R$-site.
\begin{figure}[htbp]
 \begin{center}
  \includegraphics[width=150mm]{fig1.eps}
 \end{center}
 \caption{Temperature dependence of magnetization for (a) Y$_{1-y}$La$_y$BaMn$_2$O$_6$ (0 \(\leq\) $y$ \(\leq\) 0.45) and (c) 0.45 \(\leq\) $y$ \(\leq\) 1\@.
Temperature dependence of resistivity for (b) 0 \(\leq\) $y$ \(\leq\) 0.45 and\\ 
(d) 0.45 \(\leq\) $y$ \(\leq\) 1 in a zero field (solid symbols) and magnetic fields (80 kOe, open symbols)\@.
The inset of (b) shows temperature dependence of {\rm d}log$\rho$/{\rm d}$T$ of $y$ = 0.4 and 0.45, Arrows indicate the COI transition temperature\@.}
\label{fig:one}
\end{figure}
\section{Experiment}
$A$-site ordered Y$_{1-y}$La$_y$BaMn$_2$O$_6$ were prepared in a polycrystalline form by solid state reaction.
Mixed powders of Y$_2$O$_3$, La$_2$O$_3$, BaCO$_3$, and Mn$_3$O$_4$ with a prescribed ratio were sintered at 1653-1673~K in Ar with a few intermediate grindings, and then annealed at 873~K in O$_2$.
Structural analysis was carried out by powder X-ray diffraction (XRD) method using a RIGAKU RINT-2100 diffractometer with Cu $K\alpha$ radiation at room temperature.
We confirmed that all synthesized samples have the partially $A$-site-ordered perovskite structure as designed without impurity phase by XRD patterns.
Magnetic and transport properties were measured using a Quantum Design, Physical Property Measurement System (PPMS) from 5 to 400~K.

\section{Results and discussion}
We show the temperature dependence of the magnetization and resistivity of Y$_{1-y}$La$_y$BaMn$_2$O$_6$ (0 \(\leq\) $y$ \(\leq\) 0.45) in Figs.\ 1(a) and (b)\@.
YBaMn$_2$O$_6$ ($y$ = 0) undergoes COI transition at $T_{\rm CO}$~=~500~K and AFM transition at $T_{\rm N}$ = 190~K, which is consistent with the previous report[5]\@.  
The magnetization of $y$ = 0.4 shows a slight increase below \(\sim\)150~K\@. 
With an increase in $y$ from 0.4, the ferromagnetic correlation is steeply developing.
As shown in the inset of Fig.1(b), {\rm d}log$\rho$/{\rm d}$T$ of $y$ = 0.4 and 0.45 show clear anomalies at 390~K and 360~K, respectively, below which the resistivities of $y$ = 0.4 and 0.45 are insulating.
\begin{figure}[htbp]
 \begin{center}
  \includegraphics[width=150mm]{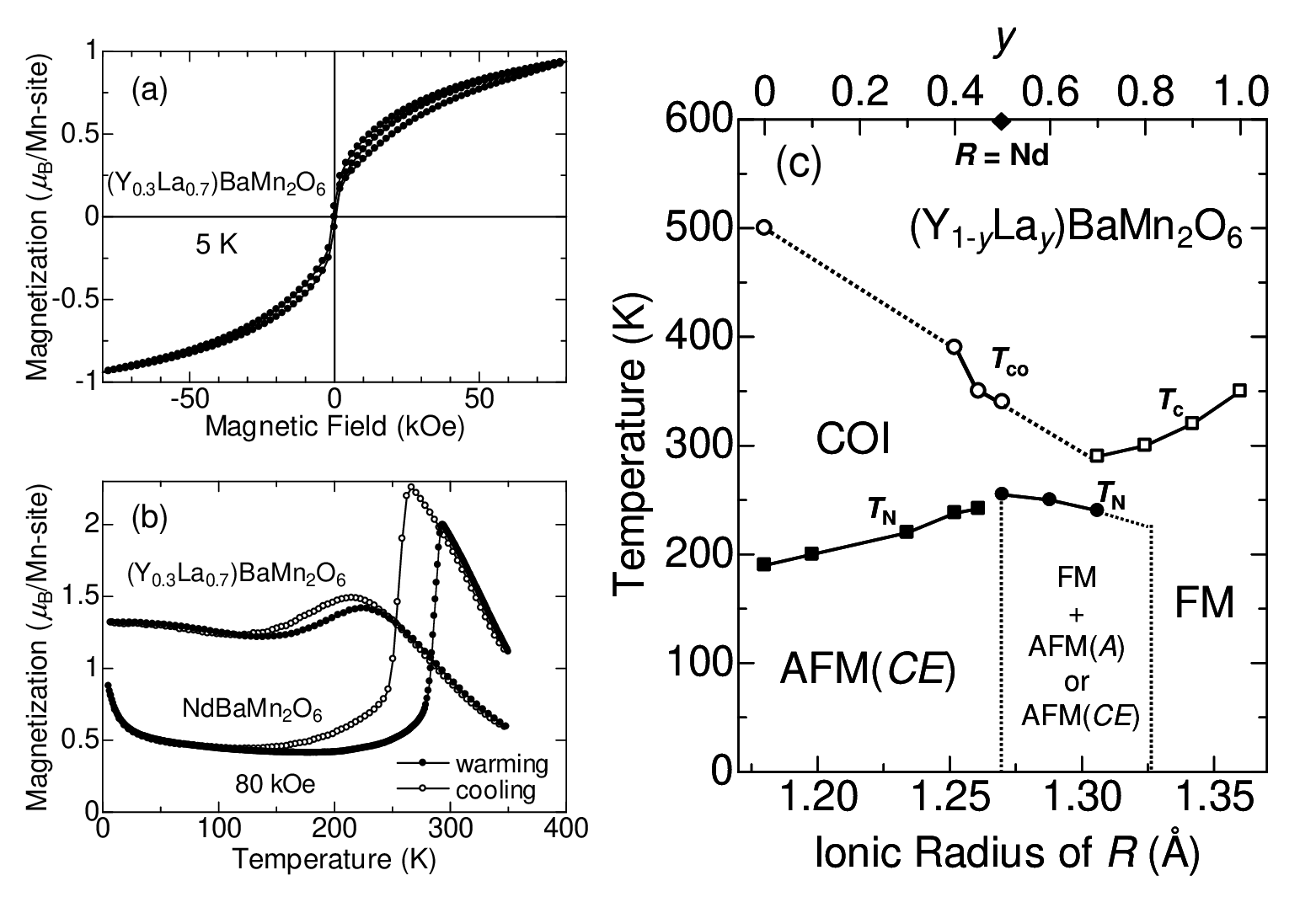}
 \end{center}
 \caption{ (a) Magnetic-field dependence of magnetization of $y$ = 0.7 at 5~K\@.
(b) Temperature dependence of magnetization for $y$ = 0.7 and NdBaMn$_2$O$_6$ (for comparison) in a field of 80 kOe.
Open and solid symbols mean the cooling and warming runs, respectively.
(c) The electronic phase diagrams for $A$-site ordered Y$_{1-y}$La$_y$BaMn$_2$O$_6$.
FM: ferromagnetic metal, AFM($A$): $A$-type antiferromagnetism, AFM($CE$): $CE$-type antiferromagnetism[3], COI: charge/orbital ordered insulator.
$T_{\rm C}$, $T_{\rm CO}$, and $T_{\rm N}$ denote ferromagnetic, charge/orbital ordering, and antiferromagnetic transition temperatures, respectively.}
 \label{fig:one}
\end{figure}
The anomaly is a piece of evidence for the occurrence of the COI transition.
The appreciable MR is not observed in Y$_{1-y}$La$_y$BaMn$_2$O$_6$ (0 \(\leq\) $y$ \(\leq\) 0.45), indicating that the COI phase is robust against external magnetic fields.

The magnetic and transport properties of Y$_{1-y}$La$_y$BaMn$_2$O$_6$ (0.45 \(\leq\) $y$ \(\leq\) 1) are shown in Figs.~1(c) and (d), respectively.
Further increasing in $y$ from 0.45 not only enhances the ferromagnetic correlation but also largely drops  the resistivity.
No distinct anomaly due to the COI transition is found in the resistivity of 0.6 \(\leq\) $y$\@.
Figure 2(a) presents the magnetization curve of $y$ = 0.7 at 5~K\@.
Ferromagnetic behavior is observed in the magnetization curve, but the ferromagnetic moment is much lower than the expected value of 3.5{\(\mu\)$_B$}/Mn-site.
The temperature dependence of the magnetization of $y$ = 0.7 and NdBaMn$_2$O$_6$ in a field of 80 kOe is shown in Fig.\ 2(b)\@. 
In $y$ = 0.7, the first-order magnetic transition accompanied by thermal hysteresis is seen around 200~K\@.
It should be noted that both the $A$-type AFM and COI transitions are of first-order, accompanying the decrease in the magnetization.
Thus, it is difficult to distinguish which transition occurs at the temperature from the magnetization alone. 
In the phase diagram for the parent compound $R$BaMn$_{2}$O$_{6}$, the $A$-type AFM state lies in the vicinity of $R$ = Nd-Pr.
The steep decrease of magnetization observed in NdBaMn$_2$O$_6$ at 250~K for cooling is due to the onset of the $A$-type AFM transition[4]\@.
Since the ionic radius of (Y$_{0.3}$La$_{0.7}$) is close to that of Pr, and the COI phase is fragile against chemical disorder, the magnetic anomaly of $y$ = 0.7 around 200~K is likely due to the $A$-type AFM transition. 
Relatively large magnetic moment of $y$ = 0.7 below $T_{\rm N}$ suggests that the FM, and $A$-type AFM (or $CE$-type) phases coexist.

Figure 2(c) displays the electronic phase diagram for {Y$_{1-y}$La$_y$BaMn$_2$O$_6$} as a function of the average ionic radius of the $R$-site, or equivalently the variation of $y$\@.
$T_{\rm CO}$ is lowered with an increase in $y$, just like the case of $A$-site ordered {\RBMO} without $R$-site randomness[5]\@.
The coexistence of the FM, and $A$-type AFM (or $CE$-type) phases is observed in 0.6 \(\leq\) $y$ \(\leq\) 0.8, i.e., near the multicritical region for the parent compound $R$BaMn$_{2}$O$_{6}$\@.
In the phase separated region, the MR is observed (Fig.\ 1(d)) and most enhanced in $y$ = 0.7\@.
In a field of 80 kOe, the resistivity of $y$~=~0.7 is reduced by about two orders of magnitude at 5~K while the MR of the end material, LaBaMn$_2$O$_6$, is much less than that of $y$ = 0.7\@.
In Y$_{1-y}$La$_y$BaMn$_2$O$_6$, the phase separation phenomenon caused by $R$-site randomness is important for the large MR effect.
The origin of the MR is explained as follows. 
By applying magnetic fields, the FM clusters are aligned along the direction of applied magnetic fields, and/or the volume of the FM clusters is increased.
As a result, the large MR is observed.

\section{Conclusion}
In summary, we have investigated the magnetic and transport properties of $A$-site ordered Y$_{1-y}$La$_y$BaMn$_2$O$_6$ to clarify the $R$-site randomness effect.
In 0 \(\leq\) $y$ \(<\) 0.6, increase in $y$ reduces $T_{\rm CO}$\@.
$R$-site (Y,La) disordering causes the phase separation.
The phase separation occurs among the FM, and $A$-type AFM (or $CE$-type) in 0.6 \(\leq\) $y$ \(\leq\) 0.8\@.
In the phase separated region, the large MR effect is found.
External magnetic fields align the FM clusters along the same direction, and/or the volume of the FM clusters is increased with increasing in magnetic fields, which largely reduce the resistivity.

\section*{Acknowledgments}
This work was partly supported by Grant-in Aid for JSPS Fellows, Scientific Research(C), the Mazda Foundation and the Asahi Glass Foundation.

\section*{References}
\numrefs{99}
\item Imada M, Fujimori A, and Tokura Y 1998 {\it Rev. Mod. Phys.} {\bf 70} 1039
\item Nakajima T, Kageyama H, and Ueda Y 2002 {\it J. Phys. Chem. Solids} {\bf 63} 913
\item Nakajima T, Kageyama H, Yoshizawa H, and Ueda Y 2002 {\it J. Phys. Soc. Jpn.} {\bf 71} 2843
\item Akahoshi D, Uchida M, Tomioka Y, Arima T, Matsui Y, and Tokura Y 2003 {\it Phys. Rev. Lett.} {\bf 90} 177203
\item Nakajima T, Yoshizawa H, and Ueda Y 2004 {\it J. Phys. Soc. Jpn.} {\bf 73} 2283
\endnumrefs

\end{document}